\documentclass[twocolumn,epsfig,prl,amsmath,amssymb]{revtex4-1}
\usepackage{graphicx}
\begin{document}
\bibliographystyle{apsrev}

\title{Gravity-analogy    in  one-dimensional ideal Fermi fluids \\
and Burgers' equation}
\author{Stefano Giovanazzi}\email{Electronic address: stevbolz@yahoo.it} 
\affiliation{Kirchhoff Institut f{\"u}r Physik, University of Heidelberg, Im Neuenheimer Feld 227, 69120 Heidelberg}
\begin{abstract}
An  hydrodynamic description of a one-dimensional flow of an ideal 
Fermi fluid is constructed from a semiclassical approximation.
For an initially fully degenerate fluid, Euler and continuity hydrodynamic equations are dual to two  
uncoupled inviscid Burgers' equations. Yet the price for the initial simplicity of the description
is paid by the complexity of non-linear instabilities towards possible turbulent evolutions.
Nevertheless, it is shown that linear long-wavelength density perturbations on a stationary flow are 
generically stable. Consequently, linear sound obeys a wave equation with analogy to gravity.
The results have applications for ultra-cold atomic gases.
\end{abstract}
\maketitle

The behavior of a quantum field in the curved space-time of a black hole 
can be modeled by sound in a convergent fluid flow. 
For this, it is sufficient to have a local correspondence of equations: 
On one side the equations governing the propagation of a quantum field near the event horizon;
On the other side those governing the sound motion near the sonic horizon of a transonic flow.
This correspondence has been first considered for a transsonic and 
irrotational inviscid-fluid flow by Unruh in his original work \cite{Unruh81}. 
The remarkable consequence of such correspondence is Unruh prediction of thermal acoustic 
radiation emitted in transonic conditions. This prediction of acoustic Hawking radiation 
(sometimes referred as sonic Hawking radiation)
suggested a way to test black hole evaporation in the laboratory.

Acoustic Hawking radiation had attracted considerable interest and
many aspects of it has been explored, such as the trans-Planckian problem 
just to cite one 
(see for instance \cite{Unruh05} and references therein).
Moreover, gravitational analogues has been considered in different systems 
 and in particular in specific quantum fluids, such as superfluids  helium-4, helium-3,
atomic Bose-Einstein condensates and one-dimensional (1D) 
Bose or Fermi degenerate fluid
(for gravitational analogues see \cite{GravitationalAnalogues,ValenciaWorkshop2009} 
and references therein).

A microscopic model of acoustic Hawking radiation that offers a way to visualize 
the source of the thermal radiation is desirable.
A good quantum fluid candidate for such a model 
is the degenerate weakly-interacting atomic Bose gas 
where the large-scale fluctuations physics is mostly understood.
This quantum fluid in a toroidal trap has indeed earlier suggested
 as a possible analogue of a sonic black hole \cite{Garay01} 
where to observe Hawking-like fluctuations effects.

An even more fundamental model where the microscopic calculations
can be done ab-initio from the many-particle Schroedinger equation is that based
on a one-dimensional (1D) ideal Fermi-degenerate fluid \cite{Giovanazzi05}.
This model is significantly more elementary than that based on the weakly-interacting
 atomic Bose gas in the corresponding one spatial dimension since both the complex
 physics of the last one and the inter-particle interaction are  absent.
Nevertheless, even if the  ab initio calculations on the 1D ideal Fermi fluid are simple to follow,
the corresponding hydrodynamic description is more tricky.
Such problematic is introduced in the present work.
It is shown here that for an initially fully degenerate fluid, the Euler and continuity 
equations are dual to two  uncoupled inviscid Burgers' equations. 
This suggests the possibility of experimentally simulating
a quantum version of Burgers' equation with ultra-cold Fermi atomic gases.
Moreover, it is show here that linear sound propagating on a stationary flow 
background is normally not evolving towards turbulent solutions in contrast to 
other solutions of the inviscid Burger equation. 
This guaranties the existence of solutions for the sound wave equation and
establishes  the analogy to gravity.

\emph{Semiclassic equations of motion}
---
The hydrodynamic equations for a Fermi-degenerate system of non-interacting particles
 are constructed starting from the lowest possible order of a semiclassical approximation.
That is of a classical dynamical system of non-interacting particles described by a phase-space 
distribution $f$, which evolves in a classical way according to the Liouville equation, and where
quantum mechanics and Fermi statistics enters only in the choice of the initial distribution.
It is assumed that $f$ varies in the spatial coordinate very slowly compared to the local mean 
inter-particle separation.
We shall indicate by $f(x,p,t)$ such semiclassical distribution of the fluid particles in phase-space 
as function of time $t$, position $x$  and particle momentum $p$. 
The particle density and the total local momentum (mass current) as function of spatial and time 
coordinates and are given by
\begin{eqnarray}
n(x,t) &=& \int  f(x,p,t) \,dp 
\label{density_def}\\
\Pi(x,t) &=& \int p\, f(x,p,t) \,dp 
\label{density_momentum_def}
\end{eqnarray}
respectively.
The equation of motion for $f(x,p,t)$ is given by the Liouville equation
\begin{eqnarray}
\frac{\partial }{\partial t} f + \frac{\partial }{\partial x} f  \frac{\partial }{\partial p}  H -
 \frac{\partial }{\partial p} f  \frac{\partial }{\partial x}  H = 0
\label{Liouville_eq}
\end{eqnarray}
where $H=H(x,p,t)=p^2/2m+V_{ext}(x,t)$ is the Hamiltonian of a single particle moving in an 
external potential $V_{ext}(x,t)$.
The general solution of the Liouville equation (\ref{Liouville_eq}) is formally obtained by the 
knowledge of the general solution of the trajectories $x(x',p',t)$ and $p(x',p',t)$ that are solution 
of the Hamilton equation of motion with initial condition $\{x',p'\}$ at $t=t_0$.
Then $f(x',p',t)=f_0(x(x',p',-t),p(x',p',-t))$ is a formal solution of the above Liouville equation.
For instance, in the case of absence of external forces the solution is given by  $f(x,p,t)=f_0(x-p t/m,p)$. 
For other possible approaches to the formal solution of the Liouville equation see Reference  
\cite{LiboffBook}.

\emph{Inviscid Burgers' equation}
----
We limit here our consideration to the case of an initial fully degenerate ideal Fermi fluid.
The quantum mechanical degeneracy of such a fluid is introduced via an
initial condition $f_0$ with sharp left and right Fermi momentum $p_0^L(x)$ and $p_0^R(x)$. 
This is given by
\begin{eqnarray}
f_0(x,p)=\frac{1}{h}\Theta[p-p_0^L(x)] \Theta[p_0^R(x)-p]
\label{f0}
\end{eqnarray}
where $h$ denotes the Planck constant.
Since $f_0$ is then uniquely defined by the value of the left and right Fermi momentum $p_0^L(x)$ 
and $p_0^R(x)$ we can look for the  equations of motions for these quantities. 
Indeed, a formal solution of the Liouville equation (\ref{Liouville_eq}) with (\ref{f0}) as initial condition
is given by the distribution
\begin{eqnarray}
f(x,p,t)=\frac{1}{h}\Theta[p-p^L(x,t)] \Theta[p^R(x,t)-p]
\label{gen_sharp_sol}
\end{eqnarray}
provides  the left and right Fermi velocities $v_L(x,t)=p_L(x,t)/m$ and $v_R(x,t)=p_R(x,t)/m$
satisfy both and independently the inviscid 1D Burgers' equation  
\cite{BurgersEquationReview}
\begin{eqnarray}
\frac{\partial }{\partial t}v_L(x,t) +v_L(x,t)\frac{\partial }{\partial x}v_L(x,t)&=&- 
\frac{\partial }{\partial x} \frac {V_{ext}(x,t)}m
\label{Burgers_L_eq}\\
\frac{\partial }{\partial t}v_R(x,t) +v_R(x,t)\frac{\partial }{\partial x}v_R(x,t)&=&- 
\frac{\partial }{\partial x} \frac {V_{ext}(x,t)}m
\label{Burgers_R_eq}
\end{eqnarray}
Note however that in order (\ref{gen_sharp_sol}) to be a solutions of the Liouville 
equation (\ref{Liouville_eq}) one has to take the velocity fields $v_L(x,t)$ and $v_R(x,t)$ 
as a multi-valued function of $x$. 
This is indeed a known fact not restricted to the dissipation-less case \cite{BurgersEquationReview}.
In the following we shall assume to be close to a stationary  distribution with single-valued Fermi edges 
where infinitesimal perturbations (sound) usually remain single valued for an arbitrarily long time.

\emph{Hydrodynamic equations} 
---
We shall assume in the following that the left and right Fermi velocities $v_L(x,t)$ and $v_R(x,t)$ are 
single-valued.
The hydrodynamic equations follows by
taking the sum and difference of the left and right Burgers' equations (\ref{Burgers_L_eq}) and 
(\ref{Burgers_R_eq}). 
We obtain in this way the equation for the mean Fermi velocity 
\begin{eqnarray}
v_F(x,t) = [v_R(x,t)-v_L(x,t)]/2
\end{eqnarray}
and the equation for the mean velocity 
\begin{eqnarray}
v(x,t) = [ v_L(x,t)+v_R(x,t) ]/2
\end{eqnarray}
 given by 
\begin{eqnarray}
\frac{\partial }{\partial t}v_F(x,t) &=& -\frac{\partial }{\partial x}\left[ v(x,t) v_F(x,t) \right] \label{sum}\\
\frac{\partial }{\partial t}v(x,t) &=& -\frac{\partial }{\partial x} \left[ \frac{ v^2(x,t)}2 + \frac { v_F^2(x,t)}2 
+\frac {V_{ext}(x,t)}m \right]\;\;\;\;\;\;
\label{difference}
\end{eqnarray}
Since the fluid density (\ref{density_def}) and the momentum density (\ref{density_momentum_def}) 
are respectively
 given by 
\begin{eqnarray}
n(x,t) &=& \frac{2 m}{h} v_F(x,t)\\
\Pi(x,t) &=& \frac{2 m^2}{h} v(x,t) v_F(x,t) = m v(x,t) n(x,t)
\end{eqnarray}
we can recognized from (\ref{sum}) and (\ref{difference})
 the continuity and Euler equation, respectively given by
\begin{eqnarray}
\frac{\partial }{\partial t}n(x,t) &=& -\frac{\partial }{\partial x}\left[ v(x,t) n(x,t) \right] \\
\frac{\partial }{\partial t}v(x,t) &=& -\frac{\partial }{\partial x} \left[ \frac { v^2(x,t)}2 +  w[n(x,t)] 
+\frac { V_{ext}(x,t)}m \right]\;\;\;\;\;
\end{eqnarray}
with an enthalpy $w$ given by
\begin{eqnarray}
w(n)=\frac{h^2 n^2}{8 m^2}
\end{eqnarray}
The local value of the speed of sound is as usual then  fixed by the relationship $c^2 = n (d w/d n) $ 
and it is given by the local value of the Fermi velocity
\begin{eqnarray}
c = v_F(x,t)
\end{eqnarray}

\emph{Sound propagation and gravity wave equation}
---
In the following we linearize the hydrodynamic equation and derive the wave equation for sound in 
a stationary fluid flow
characterized by density $n_0$ and velocity $v_0$ which are both assumed to be constant in time.
We shall indicate by 
\begin{eqnarray}
\delta n (x,t)&=& n(x,t)-n_0(x)\\
\delta v (x,t)&=& v(x,t)-v_0(x)
\end{eqnarray}
the density and velocity displacement from the stationary solution of flow, respectively.
The linearized continuity and Euler equation read
\begin{eqnarray}
\frac{\partial }{\partial t} \delta  n(x,t) &=& - \frac{\partial }{\partial x}\left[ \delta v(x,t) \, n_0(x) + 
v_0(x)\, \delta n(x,t) \right] 
\label{LinearContinuityEq}\\
\frac{\partial }{\partial t} \delta v(x,t) &=& - \frac{\partial }{\partial x} \left[ v_0(x) \,\delta v(x,t) + 
\frac{c_0^2(x)}{n_0(x)} \delta n(x,t)  \right]\;\;\;\;\;
\label{LinearEulerEq}
\end{eqnarray}
where $c_0^2(x) = n_0(x)\,w'[n_0(x)] $ is  the square of the local value of the  speed of sound.
After introducing the velocity potential $\Phi(x,t) = \int^x v(x',t) dx'$ and the velocity potential displacement 
\begin{eqnarray}
\phi(x,t) = \int^x \delta v(x',t) dx'
\end{eqnarray}
and some manipulation, we obtain finally the standard wave equation for 
$\phi(x,t)$ by eliminating $ \delta n(x,t)$ in the linearized  Euler equation (\ref{LinearEulerEq}) 
using the linearized continuity equation (\ref{LinearContinuityEq})
\begin{eqnarray}\Bigl[ \frac{\partial }{\partial t} + \frac{\partial }{\partial x}v_0(x) \Bigr]\frac{n_0(x)}{c_s^2(x)}
\Bigl[ \frac{\partial }{\partial t} + v_0(x)\frac{\partial }{\partial x} \Bigr]\phi(x,t)\;\;\;\;\;\nonumber\\
=\frac{\partial }{\partial x} \Bigr[n_0(x)\frac{\partial }{\partial x} \Bigr] \phi(x,t) \;\;\;\;\;
\label{eq:wave1}
\end{eqnarray}
The above equation is the 1D version of Unruh's  wave equation \cite{Unruh81},
that is precisely the equation for a massless scalar field in a geometry with a metric given by
\begin{eqnarray}
ds^2=\frac {n(x)}{c(x)} \Bigl\{\Bigl[ c^2(x)-v_0^2(x) \Bigl] dt^2 + 2  v_0(x) dt dx - dx^2 \Bigl\} \;\;\;\;\;\;\;
\end{eqnarray}

\emph{Peculiarity of the acoustic-wave equation}
---
The derivation in the previous subsection is a standard procedure and it is valid not only for the specific case 
of a non-interacting Fermi fluid in one dimension.
What is peculiar of the non-interacting Fermi liquid in one dimension is the separation of left and 
right movers, which is explicit in the two uncoupled Burgers' equations and in their linearized counterparts.
The consequence of this is that there is not scattering.
This may sound  a tautology since the particles of the fluid are assumed to not scatter and to be non-interacting,
but it is worth to see the consequences at the level  of the hydrodynamic equations.
The wave equation (\ref{eq:wave1}) can be rewritten for the present case as
\begin{eqnarray}
\Bigl[ \frac{\partial }{\partial t} + \frac{\partial }{\partial x}v_0(x) \Bigr]\frac{1}{v_F(x)}
\Bigl[ \frac{\partial }{\partial t} + v_0(x)\frac{\partial }{\partial x} \Bigr]\phi(x,t)\;\;\;\;\;\nonumber\\
=\frac{\partial }{\partial x} \Bigr[n_0(x)\frac{\partial }{\partial x} \Bigr] \phi(x,t) \;\;\;\;\;
\label{eq:wave1Fermi}
\end{eqnarray}
The left and right movers stationary sound solution of equation (\ref{eq:wave1Fermi}) reads
\begin{eqnarray}
\phi_L(x,t)&=&\exp\Bigl[ i \omega \Bigl( t+\int^x \frac {dy} {v_F(y)-v_0(y)} \Bigl) \Bigl]
\label{gen_sol_L}\\
\phi_R(x,t)&=&\exp\Bigl[ i \omega \Bigl( t-\int^x \frac {dy} {v_F(y)+v_0(y)} \Bigl) \Bigl]
\label{gen_sol_R}
\end{eqnarray}
as one can easily check by explicitly replace them into equation (\ref{eq:wave1Fermi}), 
(and) or into the linearized Burgers' equations, which reads
\begin{eqnarray}
\frac{\partial }{\partial t} \delta v_L(x,t) &+&v_L^0(x)\frac{\partial }{\partial x} \delta v_L(x,t)
\nonumber \\ 
&+&\delta v_L(x,t)\,\frac{\partial }{\partial x}v_L^0(x) =0
\label{linear_Burgers_L_eq}\\
\frac{\partial }{\partial t} \delta v_R(x,t) &+&v_R^0(x)\frac{\partial }{\partial x} \delta v_R(x,t)
\nonumber \\ 
&+&\delta v_R(x,t)\,\frac{\partial }{\partial x} v_R^0(x) =0
\label{linear_Burgers_R_eq}
\end{eqnarray}
The reason of this absence of coupling is also related to the constant of the ratio of density and speed of sound 
that is given by
\begin{eqnarray}
\frac {n(x)}{c(x)} = \frac {2m}h 
\end{eqnarray} 
In fact, other fluids with gravity-analogy do not have the property that 
(\ref{gen_sol_L}) and (\ref{gen_sol_R}) are exact solutions of the linearized hydrodynamic equation
but are actually an approximation of it consisting in neglecting an effective potential proportional to 
spatial derivative of the so called conformal factor $\xi(x) n(x)$  
($\xi = \hbar/ m c$ being the vacuum correlation length).
This effective potential can scatter sound and it is often neglected in discussing acoustic 
Hawking radiation. The conformal factor for a 1D ideal Fermi fluid is $\xi(x) n(x)=1/\pi$.

\emph{Peculiarity of stationary flow} 
---
A different but minor peculiarity of the fluid in consideration concerns the stationary flow 
equations. Let us indicated by $I$ the conserved particle current $n_0(x) v_0(x) $. 
Then stationary equations are given by
\begin{eqnarray}
v_0^2(x)+v_F^2(x) &=& \frac {2(\mu-V_{ext})}m \\
v_0(x)v_F(x) &=& \frac {h}{2 m} I 
\end{eqnarray}
where $\mu$ is an integration constant.
Note the symmetry in which the flow velocity and the Fermi velocity
appear in the above equations.
Expressing the above equations using the left and right Fermi velocities
\begin{eqnarray}
v_R(x) = v_F(x)+v_0(x)\\
v_L(x) = v_F(x)-v_0(x)
\end{eqnarray}
 which represent  the local edges of the velocity distribution, we obtain two uncouple equation
\begin{eqnarray}
\frac 12 m \,v_R^2(x) &=& \mu_R-V_{ext}(x)  \\
\frac 12 m \,v_L^2(x) &=& \mu_L-V_{ext} (x) \label{stationaryeq}
\end{eqnarray}
These equations are of course just the energy conservation for the left and right Fermi edge,
where the left and right chemical potential are given respectively by
\begin{eqnarray}
\mu_R &=& \mu   + \frac 12 h I \\
\mu_L &=& \mu   - \frac 12 h I 
\end{eqnarray}
The metric written in terms of the left and right Fermi velocities is given by
\begin{eqnarray}
ds^2= \frac {2m} h \Bigl[ v_R \, v_L \, dt^2 + ( v_R + v_L ) \, dt dx - dx^2 \Bigl] \;\;\;\;\;\;\;\;\;
\end{eqnarray}

\emph{Event horizon}
---
Now we consider the particular case of a sonic event horizon. 
Let be $v_L$ the downstream mode that has an horizon in $x=0$. 
This is realized by taking $V_{\text{ext}}(x) = V_{\text{max}} - \frac 12 m \kappa^2 x^2$ 
as external potential and $\mu_L  = V_{\text{max}} $.
By inserting these in equation (\ref{stationaryeq}) we obtain the equation 
for the left Fermi velocity in presence of the horizon
\begin{eqnarray}
\frac{\partial }{\partial t}v_L(x,t) +\frac{\partial }{\partial x} \left\{ \frac 12 v_L^2(x,t) 
 - \frac 12 \kappa^2 x^2 \right\} &=& 0
\end{eqnarray}
A stationary solution for the left moving edge which is differentiable is given by 
\begin{eqnarray}
v_L^{(0)}=\kappa x
\end{eqnarray}
The equation for a displacement of the left Fermi point $v_L=\delta v_L+ v_L^{(0)}$ is given by 
\begin{eqnarray}
\frac{\partial }{\partial t}\delta v_L(x,t) +\frac{\partial }{\partial x} \left[  \kappa x \delta v_L(x,t)  \right] &=& 0
\end{eqnarray}
and has the solution given by
\begin{eqnarray}
\delta v_L(x,t) = i \frac \omega {\kappa x}   \exp[ - i \omega t + i (\omega/\kappa) \ln(x/a)]
\end{eqnarray}

\emph{Breaking time and quantum diffusion} ---
The inviscid Burgers' equation is often used to exemplify the formation of shocks and
of multivalued solutions. The most simple example of this is that of an initial plane wave 
$\delta v(x,0)=\delta v_0 \sin(q x)$ in the absence of external forces.
The solution with this single-valued initial boundary condition
evolves into a multivalued solution after the breaking time $T_c=1/q \delta v_0$. 
To observe such non-linear effect in the 1D ideal Fermi fluid 
$T_c$ should at least be shorter that the "lifetime" associated with the linewidth of the 
elementary excitation which is of order of $\tau =2 m/\hbar q^2$.
So it is natural to ask how much such mode with wavevector $q$ 
should be populated in order that 
the non-linear effect starts to be comparable with the lifetime $\tau$. 
The answer is about one atom per half a wavelength.
In fact, indicating $\Delta N = \int_0^{\lambda /2} \delta n(x) dx= 2 m \delta v_0 / h q  $ results 
that $T_c=2m/h q^2 \Delta N= 2\pi \tau /\Delta N$.
Like viscosity in Burgers' equation tends to counteract the formation of shocks
so acts the quantum mechanical uncertainty between position and momentum
in the 1D ideal Fermi fluid.
This reminds us  the {\it viscosity bound conjecture} of P.~K.~Kovtun, D.~T.~Son, and A.~O.~Starinets
\cite{Kovtun05}.

\emph{Conclusion} ---
It is shown here that 1D ideal Fermi fluids in smooth external potentials 
admit a linearized hydrodynamic description
for sound with a wave equation with analogy to gravity.
Thus ultra-cold atomic Fermi degenerate gases of negligible interparticle interaction
in elongated traps with high radial confinement
can be used as well as other quantum 1D fluids to test 
in the laboratory the acoustic analogue of Hawking radiation.
Moreover, a typical non-linear effect in hydrodynamics, i.e.
the initial evolution towards the formation of shocks
may be experimentally studied.


\begin{thebibliography}{15}

\bibitem{Unruh81}
{W.~G.~Unruh, Phys.\ Rev.\ Lett.\ {\bf 46}, 1351 (1981).}

\bibitem{Unruh05}
{William G. Unruh and Ralf SchŸtzhold, Phys.\ Rev.\ D  {\bf 71}, 024028 (2005).}

\bibitem{GravitationalAnalogues}
{{\it Quantum Analogues: From Phase Transitions to Black Holes and Cosmology}, 
edited by R.~Sch\"utzhold and W.~G.~Unruh, Springer Lecture Notes in Physics Vol. {\bf 718} 
(Springer, Berlin Heidelberg, 2007); 
C.~Barcelo, S.~Liberati, and M.~Visser, Living Rev. Relativity {\bf 8}, 12 (2005); 
G.~E.~Volovik, {\it Universe in a Helium Droplet} (Oxford University Press, Oxford, 2003); 
{\it Artificial Black Holes}, edited by M.~Novello, M.~Visser, and G.~E.~Volovik 
(World Scientific, Singapore, 2002).}

\bibitem{ValenciaWorkshop2009}
{Workshop EHR: {\it Towards the observation of Hawking radiation in condensed matter systems} (2009), 
URL http://www.uv.es/workshopEHR.}

\bibitem{Garay01}
{L.~J.~Garay, J.~R.~Anglin, J.~I.~Cirac and P.~Zoller,  Phys.\ Rev.\ Lett.\  {\bf 85}, 4643 (2000); 
Phys.\ Rev.\ A {\bf 63}, 023611 (2001).}

\bibitem{Giovanazzi05}
{S.~Giovanazzi, Phys.\ Rev.\ Lett.\ {\bf 94}, 061302 (2005).}

\bibitem{BurgersEquationReview}
{J.~Bec, K.~Khanin, Phys.\ Rep.\ {\bf 447}, 1 (2007).}

\bibitem{LiboffBook}  
{\it Kinetic theory: classical, quantum, and relativistic descriptions}, edited by R.~L.~Liboff, 
(Prentice-Hall, New Jersey, 1990).

\bibitem{Kovtun05}
{P.~K.~Kovtun, D.~T.~Son, and A.~O.~Starinets, Phys.\ Rev.\ Lett.\ {\bf 94}, 111601 (2005).}


\end{thebibliography}
\end{document}